# Hurricane-blackout-heatwave

## Compound Hazard Risk and Resilience in a Changing Climate

Kairui Feng[a], Ouyang Min[b], Ning Lin[a,1]

[a]Civil and Environmental Engineering, Princeton University

[b] School of Artificial Intelligence and Automation,

Huazhong University of Science and Technology

**Abstract:**

Hurricanes have caused power outages and blackouts, affecting millions of customers and inducing severe social and economic impacts. The impacts of hurricane-caused blackouts may worsen due to increased heat extremes and possibly increased hurricanes under climate change. We apply hurricane and heatwave projections with power outage and recovery process analysis to investigate how the emerging hurricane-blackout-heatwave compound hazard may vary in a changing climate, for Harris County in Texas (including major part of Houston City) as an example. We find that, under the high-emissions scenario RCP8.5, the expected percent of customers experiencing at least one longer-than-5-day hurricane-induced power outage in a 20-year period would increase significantly from 14% at the end of the 20[th] century to 44% at the end of the 21[st] century in Harris County. The expected percent of customers who may experience at least one longer-than-5-day heatwave without power (to provide air conditioning) would increase alarmingly, from 0.8% to 15.5%. These increases of risk may be largely avoided if the climate is well controlled under the stringent mitigation scenario RCP2.6. We also reveal that a moderate enhancement of critical sectors of the distribution network can significantly improve the resilience of the entire power grid and mitigate the risk of the future compound hazard. Together these findings suggest that, in addition to climate mitigation, climate adaptation actions are urgently needed to improve the resilience of coastal power systems.

---

1 To whom correspondence should be addressed. Email: nlin@princeton.edu

**Main**

As forms and effects of extreme weather, hurricanes (generally called tropical cyclones or TCs) threaten 59.6 million people in the U.S. (2018) [1,2] and are important initiating causes of large-scale failures of power systems. Hurricane Maria (2017) devastated Puerto Rico's power grid, resulting in a power loss of 3.4 billion customer hours and the worst blackout in US history. Hurricane Irma (2017) deprived over 7 million customers of electricity, 2.1 million of whom still lacked access to electricity after four days [3]. Hurricane Harvey (2017), making landfall on the Texas coast, disrupted hundreds of overhead electricity lines and took away over 10,000 megawatts (MW) of electricity generation capacity; local utility companies spent over a week to restore the system to normal [4]. Hurricane Florence (2018) cut electric power for around 1.4 million customers; the system took two weeks to recover [5]. Similarly, Hurricane Sandy (2012) affected over 8.5 million customers [6,7]. These disruptions, leaving millions of customers without electricity for days, call for investigation of power system resilience and ultimately a re-design of the energy infrastructure [8]. Moreover, the U.S. power grid will become more vulnerable to weather and climate-induced failures in the future due to climate change [9]. In particular, hurricane-induced power outages are likely to become more severe, as increasing evidence shows that TC intensity will increase due to climate change (e.g., 11-16). This potential change should be quantified and accounted for in planning future energy infrastructure.

Projected with greater confidence, climate change may also induce more heat extremes [17,18]. Heatwaves are the primary cause of weather-related mortality (due to heat cramps, syncope, exhaustion, stroke, etc.) in the U.S. [19], and they may harm mental health [20,21], sleep quality [22], and social stability [23] in different ways. Ref. [24] is the first to connect hurricanes with heatwave impacts. With hurricane-induced power outages, the impact of heatwaves would dramatically increase, as air conditioning, with around 1.6 billion units in operation over the U.S. [25], is critical in reducing the vulnerability to extreme heat (the fatality risk under heatwaves without air conditioning is estimated to triple, as revealed by a post-heatwave analysis in Chicago [26]). Ref. [24] found that TC-heatwave compound events have been rare and affected only about 1000 people worldwide, as the seasonal peak of extreme heat precedes that of major TCs. Due to global warming, Ref. [24] projected that the TC-heatwave compound hazard would affect a much larger population in the future, e.g., over two million in a world 2°C warmer than pre-industrial times. However, Ref. [24]'s projection neglected potential changes in TC climatology in the future, which may result in an underestimation of the compound hazard, given that hurricanes will likely become stronger (e.g., 11-16) and possibly occur earlier in the season (as the first two power-destructive U.S. hurricanes in this year of 2020, Hurricane Luara is the earliest "L" named storm to ever form in the Atlantic and Hurricane Isaias is the earliest fifth named storm to make landfall in the U.S.). On the other hand, without coupling climate hazard analysis and power system modeling, ref. [24]'s analysis did not account for the reliability and resilience of the power system, which may result in an overestimation of the compound risk.

To better quantify the evolving joint impact of hurricanes and heatwaves, here, for the first time, we model the hurricane-blackout-heatwave compound hazard risk in a changing climate. We combine statistical-deterministically-downscaled hurricane climatology and directly-projected heatwave climatology from general circulation models (GCMs), and we explicitly model the power system failure and recovery process under the climate hazard scenarios. We investigate how the risk of local residents experiencing prolonged hurricane-blackout-heatwave compound hazards may change from the current to future climates.

To illustrate, we apply the analysis to Harris County (including major part of Houston City) in Texas (see Methods and Supplementary Fig. S1 for a description of the area's geographics and power network). As hurricanes cannot be well resolved in typical GCMs due to their relatively small scales, we apply large datasets of synthetic storms generated by a deterministic-statistical hurricane model [27] for the area [28], for the historical climate of 1981–2000 based on the National Centers for Environmental Prediction (NCEP) reanalysis and for the future climate of 2081-2100 under the emissions scenario RCP8.5 based on 6 GCMs in the fifth Coupled Model Inter-comparison Project (CMIP5; see Methods). Based on the performance of the GCMs in terms of their historical simulations compared to the reanalysis-based simulation, we bias-correct the GCM-simulated storm frequency and landfall intensity distribution and combine the 6 simulations into a single projection for the future climate (see Methods). Then we stochastically resample the synthetic storms from the combined projection to form 10,000 20-year simulations for the historical climate and 10,000 20-year simulations for the future climate. Here we focus on wind effects and apply a parametric model to estimate the spatial-temporal wind field for each synthetic storm (see Methods). We apply a physics-based power outage and recovery process model [29], evaluated with two historical cases (Hurricanes Harvey and Ike) for the study area, to simulate the wind-induced power system failure and recovery process for each synthetic storm (see Methods).

The US National Weather Service issues warnings when a forecast heat index (HI) characterizing humid heat exceeds 40.6°C. Ref. [24] defines a compound TC-heat event as a major hurricane followed within 30 days by an HI greater than 40.6°C at the site of landfall. Here we also define a heat event as an HI over 40.6°C, but we are interested in a range of time scales, especially over 5 days following the hurricane landfall. (Considering that hurricanes usually start to interrupt the life pace of local residents at least 2 days ahead of landfall [30], a greater-than-5-day power outage plus heatwave means affected residents cannot resume normal life for over a week, which is usually a benchmark for resilience design for critical infrastructure systems [31].) Based on the same reanalysis and GCM datasets, we calculate and bias-correct the HI for the study area during and after the landfall of each synthetic storm (see Methods). In addition, we modify the HI to statistically account for the interdependence between hurricanes and heatwaves (see Methods). Combining the obtained HI dataset and hurricane power outage analysis enables us to estimate the risk of Harris residents experiencing prolonged (e.g., 5 day) heatwaves (HI>40.6°C) after losing power due to hurricane wind damage. We first focus on the estimated risk for the high-emissions scenario RCP8.5 and the end-of-the-century (2081-2020) time frame (Results); we then investigate the sensitivity of the estimated risk to the emission scenario, the mid-of-the-century time frame, and, considering the large uncertainty regarding how TC frequency will change (e.g., 14,15,32,33), the storm frequency projection (Discussion and Supplementary Materials).

In addition to quantifying the risk, we explore efficient strategies to increase power system resilience for combating future hurricane-blackout-heatwave compound hazards. Consistent with previous results [7], our network analysis (see Methods) shows that local power failures have a disproportionally large non-local impact on the power system; the reliability of local power distribution networks (i.e., the final stage of energy infrastructure) is particularly critical. Thus, we propose a greedy undergrounding strategy (burying parts of the power network [34-36], see Methods) to protect a small portion of wires that are close to the root nodes of the power distribution networks. We show that such a targeted undergrounding strategy is much more efficient than the generally-applied uniform undergrounding strategy in increasing the resilience of the power system and decreasing the risk of future hurricane-blackout-heatwave compound hazards.

## Results

**Historical Cases.** Hurricanes Harvey and Ike are the main events in the past 20 years that led to significant power outages (over 10% of customers lost power) in Harris County. Ike had a higher gust wind observation (~92 mph) when it hit Houston; Harvey was weaker (<50 mph) but lasted longer and brought heavy rainfall [37]. The wind during Ike broke many more poles, leading to a larger power outage. Results from the power grid outage and recovery process model compare relatively well with the data for Hurricanes Ike and Harvey, indicating the model's success in capturing both large and relatively small power outage events (Fig. 1). The model estimates a relatively small power outage (1% [0%-1%]) after the 5-day restoration period for Harvey and a large power outage (63% [60%- 66%]) for Ike for over 5 days, agreeing with observations (Fig. 1).

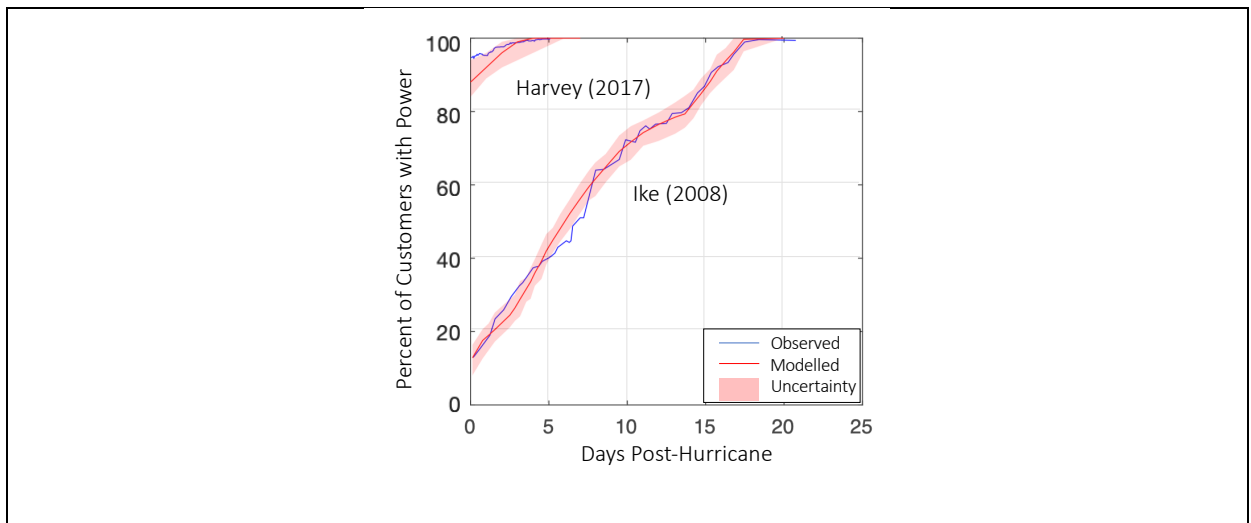

Figure 1. Simulated (red curve showing median values, with 32% to 68% quantile range shown by shade) compared with observed (blue curve) power outage and recovery process for Hurricanes Harvey and Ike in Harris County, TX.

**Post-hurricane Heatwaves.** While heatwaves did not follow Hurricanes Ike and Harvey [51], they may become more likely to follow future hurricanes. Under RCP8.5, the projected change in global mean surface air temperature for the late 21$^{st}$ century relative to the reference period of 1986-2005 is 3.7 °C based on all GCMs in the IPCC report [38]. The 6 GCMs used in this study predict an increase of the average global temperature by 3.4°C in the future climate (2081-2100) compared to that in the historical climate (1981-2000). To investigate the likelihood and duration of future post-hurricane heatwaves, we analyze the synthetic hurricane and heatwave datasets. Fig. 2a shows how the probability for a post-hurricane heatwave with a certain duration changes from the historical to future climate for the study area. The probability of a heatwave, especially a long-lasting one, following a hurricane is quite small for the historical climate, which is consistent with the observation that hurricane-heatwave hazards have affected only ~1000 people worldwide during 1979–2017 [24]. However, the probability curve is much higher for the future climate. The probability for a post-hurricane heatwave lasting over 5 days is 2.7% (1.8% - 4.2%) under the historical climate but 15.2% (9.4%-24.2%) in the future climate. For a post-hurricane heatwave to last over 12 days is almost impossible in the historical climate, but a nonnegligible probability of 5.1% (4.1%~8.5%) exists for it to happen in the future climate. To better reveal the timescale of climate change impact, Fig. 2b shows the relative climate risk, defined as the probability of post-hurricane heatwaves lasting for a certain duration in the future climate divided by that in the historical climate. The relative climate risk increases sharply with the duration, reaches the peak around 13 days, and then decreases quickly. Although the probability for a 1-day heatwave following hurricanes in the future climate would be only ~3 (2-6) times larger than that in the historical climate, the probability for a 5-day heatwave would be 5 (3 ~ 9) times larger and the probability for a 13-day heatwave would be 16 (9~24) times larger. This time-scale pattern of climate change impact should be considered when developing maintenance and management strategies for urban infrastructure systems. For example, resilience criteria of these systems may be enhanced to avoid "resonance" effects of climate change. A typical recovery cycle for the power system is currently 5 days or longer [31]; the dramatic change in the climate risk may render such a time scale of recovery not resilient.

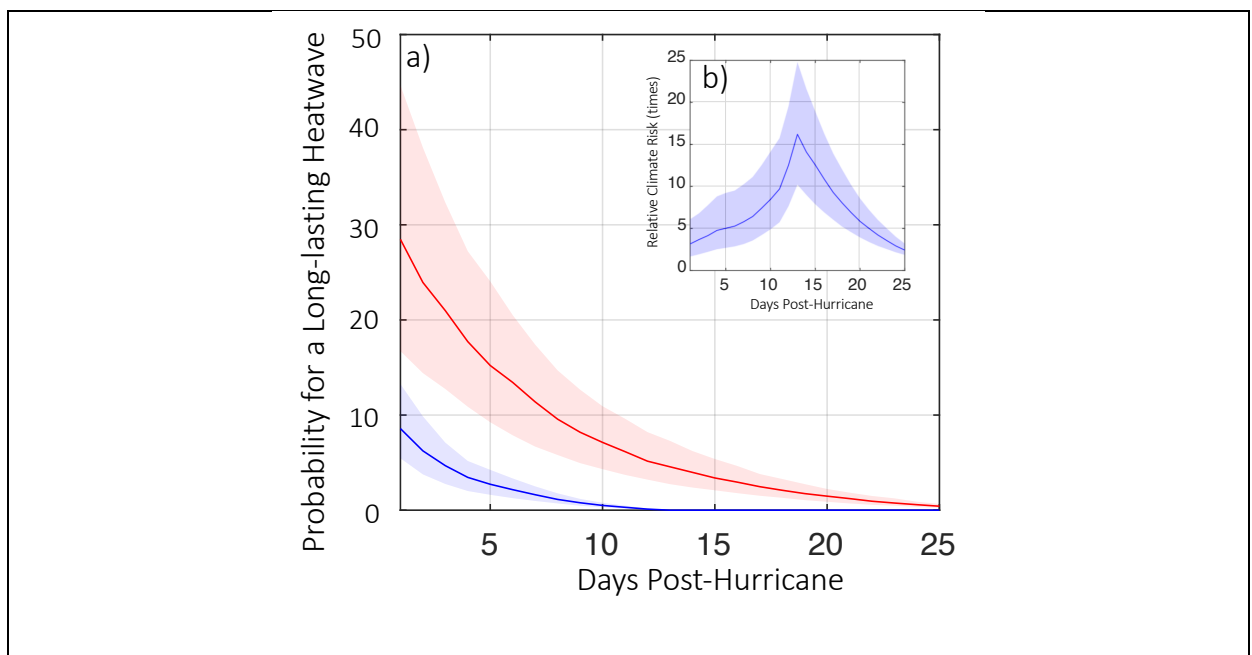

Figure 2. Analysis of likelihood and duration of post-hurricane heatwaves for Harris County, TX. a) Probability for post-hurricane heatwave with certain duration in historical climate. (blue curve showing median, with 32% to 68% quantile range shown by shade) and future climate (red curve, with 32% to 68% quantile range shown by shade). b) Relative climate risk, defined as ratio of future probability of post-hurricane heatwave and historical probability (lower-bounded to avoid numerical problems), as function of post-hurricane heatwave duration (shade shows 32% to 68% quantile range).

**Blackout and Compound Hazards.** Incorporating the power outage and recovery modeling, we analyze the hurricane-blackout and hurricane-blackout-heatwave compound hazards for the study area. Figs. 3a and 3b show the estimated percentage of Harris customers who may not experience post-hurricane power outages during a 20-year period under the historical and future climates, respectively. On average, the hurricane-induced power outage would affect 50% of customers during a 20-year period in the historical climate and 73% of customers in the future. 15% customers would not be subject to any post-hurricane blackout during the 20-year period in the historical climate, while only 5% customers would not be affected in the future. In the historical climate over a 20-year period, on average 14% of Harris residents would face at least one longer-than-5-day post-hurricane power outage, which is less than one third of the expected level of 44% in the future. For 90% of cases under the historical climate, the utility company (CenterPoint Energy) could fully repair the power system within about 15 days, which matches recent-year records (e.g., 12 days for Hurricane Irma in 2017, and 13 days for Hurricanes Michael and Florence in 2018 [39]). With the same response strategies and resources, the utility company might spend over 25 days to repair the power system in the future under severe hurricanes (10th percentile). By comparison (with Fig. 1), the probability of experiencing the scale and duration of power outage induced by Hurricane Ike during a 20-year period is about 10% in the historical climate and 35% in the future. Hurricane Harvey is an under-average event in both the historical (87%) and future (96%) climates. These analyses show that climate change may dramatically increase the failure risk of the power system, especially the tail of the risk that people will face.

Figs. 3c and 3d show the estimated percentage of Harris customers who may not experience any hurricane-blackout-heatwave compound hazard during a 20-year period under the historical and future climates, respectively. The chance for a customer to experience a longer-than-5-day compound hazard is almost zero (0.8%) under the historical climate. However, the chance for a customer to be affected by such a compound hazard in the future climate (15.5%) would be ~20 times larger. Recall that, due to climate change, the 5-day post-hurricane power outage risk is estimated to increase by 3 times (Figs. 3a & 3b); 5-day post-hurricane heatwave risk is estimated to increase by 5 times (Fig. 2b). The dramatic 20-time increase of hurricane-blackout-heatwave compound hazard, which is larger than simply multiplying the two factors together as if hurricanes and heatwaves are climatologically uncorrelated (15 times), indicates that large hurricanes and long-lasting heatwaves will be more likely to co-occur in the future. In other words, larger, more frequent hurricanes leading to vaster power outages will be followed by longer-lasting heatwaves, affecting over 15% of Harris residents towards the end of the century.

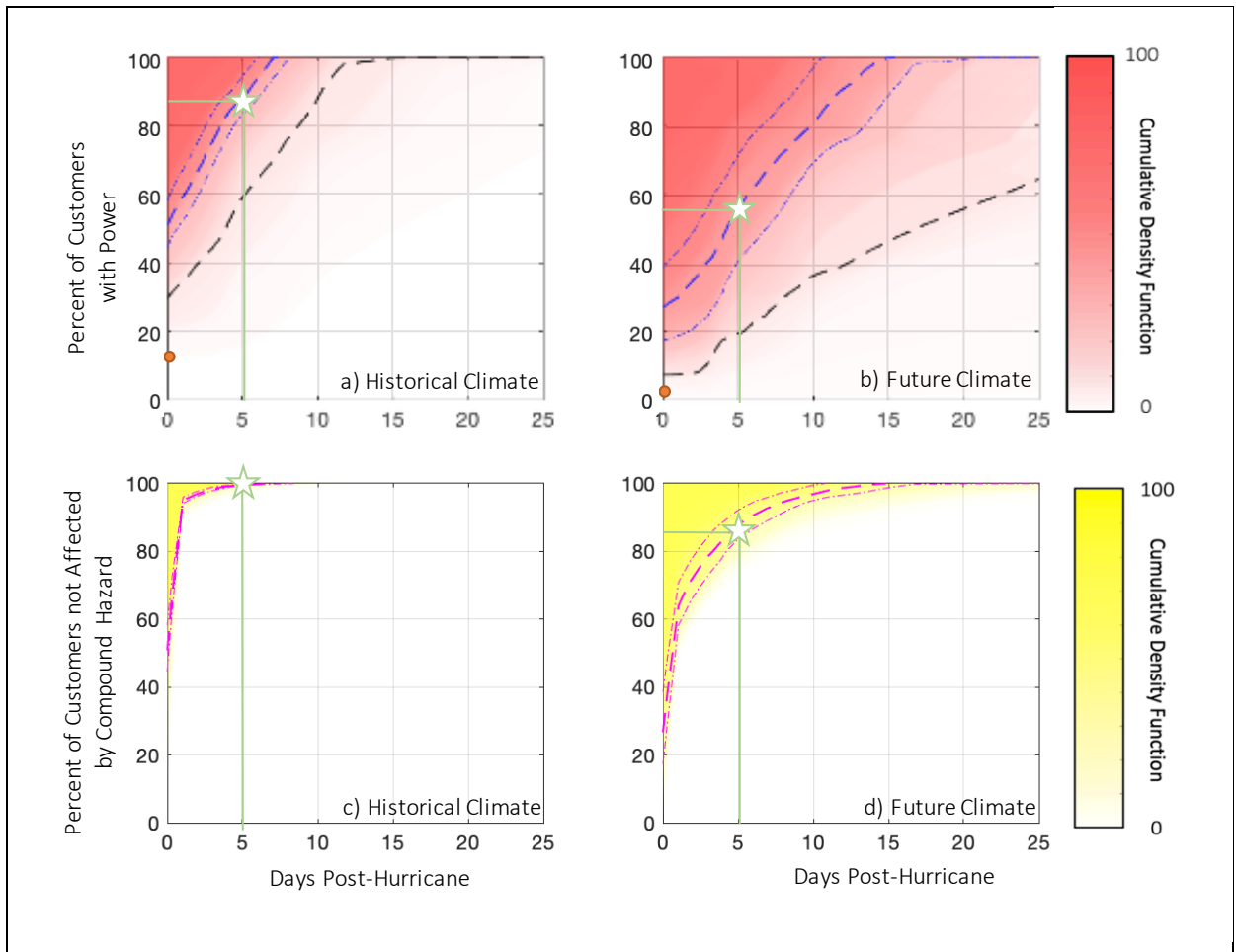

Figure 3. Analysis of hurricane-blackout-heatwave compound events for Harris County, TX. Estimated percentage of customers with power days after hurricane landfall in a 20-year period are shown for a) historical climate and b) future climate. Orange dot shows simulated worst case of power outage over all simulations. Dashed black line shows 10% worst case of power recovery. Estimated percentage of customers not experiencing hurricane-blackout-heatwave compound hazard days after hurricane landfall in a 20-year period are shown for c) historical and d) future climate. In all panels, the dashed and dashed-dotted curves show expectation and ±1σ range based on all simulations. Shade shows CDF, indicating probability of less than certain (y) percentage of customers unaffected, with darker color corresponding to higher probability. Green stars highlight expected percent of customers not affected 5 days post hurricane landfall.

To investigate the spatial distribution of the risk, we estimate the percent of customers to experience at least one longer-than-5-day hurricane-induced power outage (Figs. 4a and 4b) and hurricane-blackout-heatwave compound hazard (Figs. 4c and 4d) for each census tract in Harris County in the historical and future climates. Changing from the historical to the future climate, the power outage risk of all the census tracts would at least double. The hurricane-blackout-heatwave compound hazard risk would increase even more dramatically. Over 95% people in all census tracts would experience no compound hazards in the historical climate; in the future, over 95% of census tracts would have over 5% of the population experiencing the compound hazard. The distribution of risk is heterogeneous, which implies a heterogeneous distribution of the resilience level of the power system, as the hurricane and heatwave hazards vary only slightly over this relatively small region. The spatial pattern shows that residents near the center of Houston (the middle and lower part of the County) may experience lower power outage and compound hazard risks than residents in rural (e.g., the upper part) areas of the County. The varying densities of power substations and spatial patterns of distribution networks induce most of this difference.

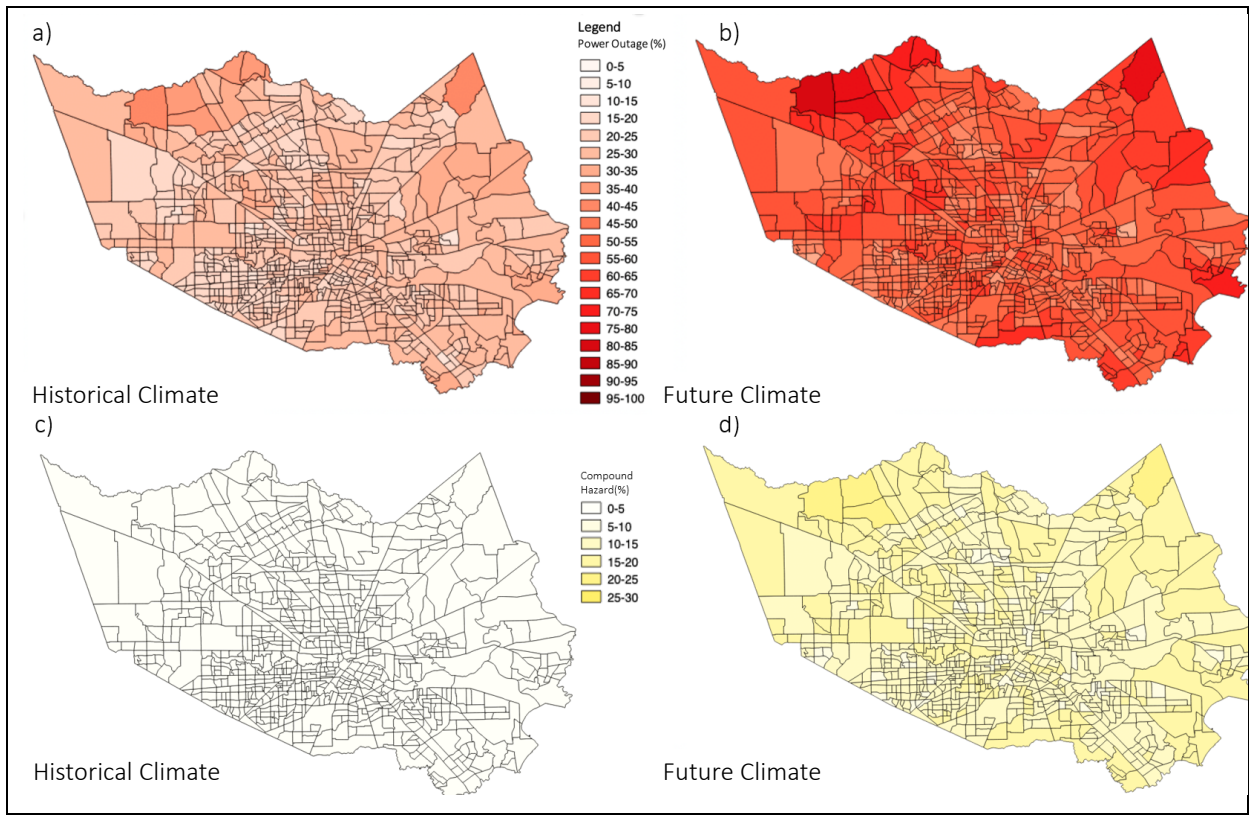

Figure 4. Estimated percentage of customers facing at least one longer-than-5-day power outage (a & b) and blackout-heatwave compound hazard (c & d) following a hurricane in historical (a & c) and future climate (b & d) for each census tract in Harris County.

**Scaling Law of Power System Failure.** To better understand the power system, we apply network analysis to investigate links between local disruptions (the failure of a specific pole or distribution/transmission line) and global failures, using our large synthetic hurricane dataset. In each event, we calculate the probability W(x) for a customer to be subject to a disruption that affect more than x customers and the probability P(x) for a disruption to affect more than x customers. W(x) also represents the percent of service interruptions induced by all local disruptions that affect more than x customers. Fig. 5a shows the obtained generalized scaling law [7] between W(x) and P(x), for each simulated event and averaged over all the events for the historical climate. Rather than a linear relationship indicating a uniform distribution of damage, the concave scaling curve shows that on average the largest 20% of local disruptions are responsible for 72% (71%~75%) of the global power outage (measured by the number of affected customers). This result obtained from the synthetic analysis for Harris County is comparable with a previous empirical result for Upstate New York that during Hurricane Sandy the top 20% of local disruptions induced 79%-89% of the service interruptions [7]. In fact, the difference in the scaling curves among the full range of synthetic hurricanes is quite small, which confirms the robustness and generalizability of the scaling law in describing a hurricane-damaged power system. Further, ref. [7] found that the scaling curve developed using data on failures in daily operation is also similar, suggesting that the power network vulnerability (i.e., local disruptions inducing large-scale interruptions) exists independently of exogenous effects. Thus, reliability-based redesign against various hurricanes may not need to differ fundamentally from the daily-operation-based enhancement.

Large power outages are usually induced by the breaking down of local damages (as shown in our analysis and previous studies [7]). Large portions of local damages are due to distribution network failures, so we analyze the connection between power outages and local distribution network topology. Fig 5b. shows the correlation between the percentage of customers experiencing a 5-day power outage (averaged over all simulated synthetic storms for the historical climate) and the harmonic mean length of the power distribution network sectors at the census tract level. The results show that the longer the length of the local power distribution networks, the higher the risk. The mean length of the power distribution networks is highly related to the pattern of urban development. A region with a lower population density may have a larger mean length of power distribution networks, as it may have fewer substations to support customers who live relatively far away. Thus, our results indicate that, for hurricane-prone regions, the scaling of urban development may have contributed significantly to the spatial distribution of power outage risks. The observed high correlation between the power outage rate and the mean length of distribution networks and the generalized scaling relationship between the probability of local failure and the failure impact, shown in Fig. 5, can be explained theoretically and generally for acyclic (the most common) power distribution networks (see theoretical analysis and Fig. S2 in Supplementary Materials).

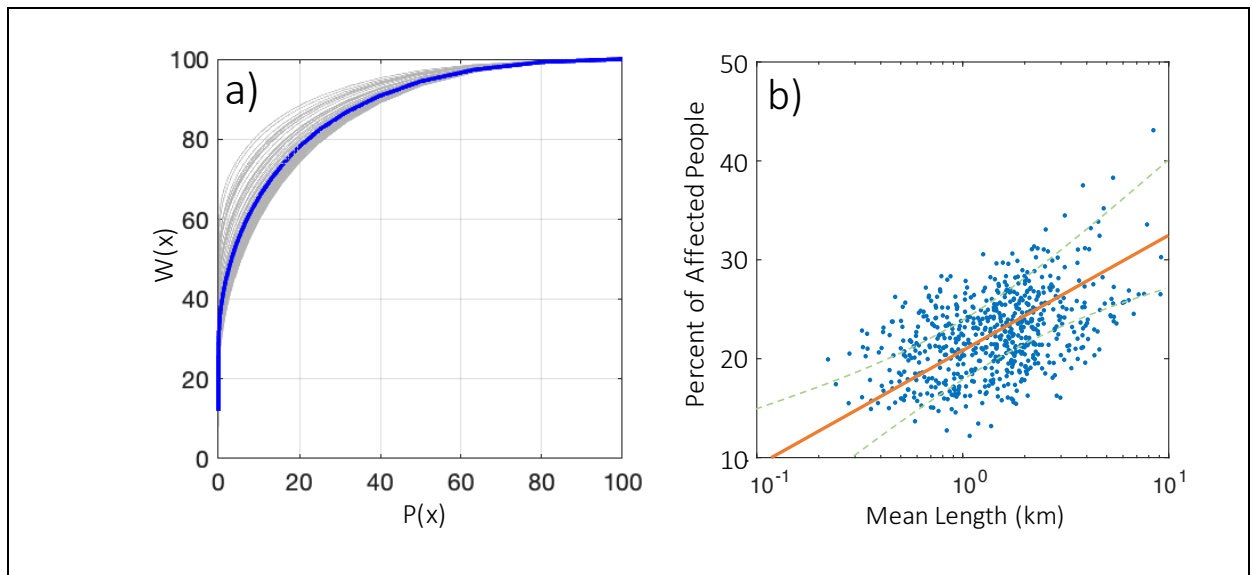

Figure 5. Network analysis of the synthetic-hurricane-induced power outage in Harris County. a) Generalized scaling law for the Harris power system: empirical probability $W(x)$ of a customer being affected by a disruption that affects more than x customers vs. empirical probability $P(x)$ for a disruption to affect more than x customers during the event. Blue curve shows the average over all synthetic events in the historical simulation; gray curves show randomly selected 100 events. b) Percentage of customers experiencing a 5-day-long power

outage averaged over all synthetic hurricanes simulated for historical climate vs. harmonic mean length (reciprocal of the arithmetic mean of reciprocals) of power distribution network sectors for each census tract in Harris County. Red line shows the linear fit; dashed curves show the ±1σuncertainty range.

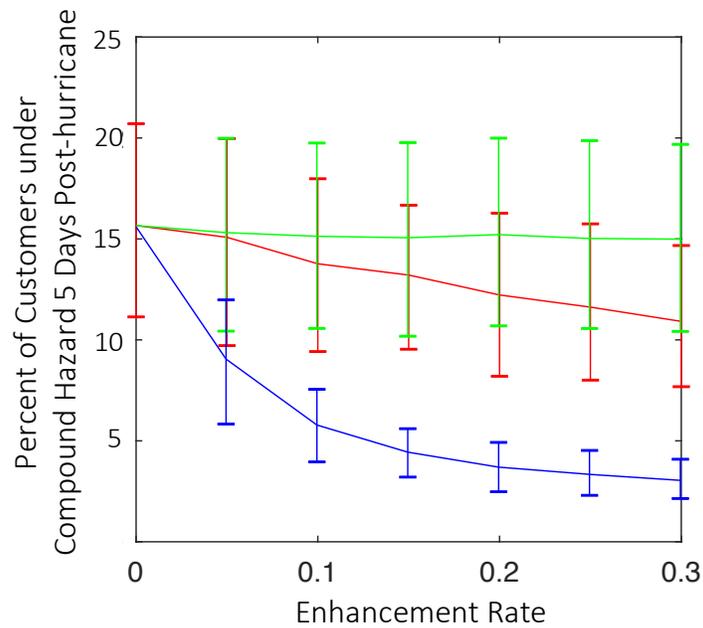

Figure 6. Impact of 5-day hurricane-blackout-heatwave compound hazard (percent of affected customers in a 20-year period) in the future as a function of network enhancement rate with three strategies: randomly undergrounding power transmission networks (green), randomly undergrounding power distribution networks (red), and greedily undergrounding power distribution networks (blue). Curves show the expectation; error bars show ±1σ uncertainty range.

**Efficient Undergrounding Strategy.** Based on an improved understanding of the power network, we test three network enhancement strategies to mitigate the hurricane-blackout-heatwave compound hazard in the future. Fig. 6 compares the reduction of the compound hazard impact as a function of the enhancement rate for the three strategies; the enhancement rate is the total length of enhanced networks divided by the total length of the network sections that could be enhanced. Random enhancement of the high-voltage transmission network (burying transmission branches randomly) provides limited improvement of the global system performance. This finding reconfirms that the local distribution networks dominate global pattern of the power outage and subsequent recovery process. Randomly undergrounding low-voltage power distribution networks improves the system performance linearly, which means that the power distribution networks without protection face the same risk as before the enhancement. Given the acyclic topology of the power distribution networks, a greedy undergrounding strategy is used to protect a small portion of wires close to the root nodes of the distribution networks. In the simulation, the algorithm simply protects the root sector of the longest overhead branches of the distribution network. The greedy undergrounding strategy improves the system performance much more than the other two strategies. For example, with the first 5% power distribution networks undergrounded, the expected percent of customers who might face the compound hazard for 5 days drops to 9.6% from 15.5% (Fig. 3d). This performance improvement (5.9%) is ~15 times more than that of randomly undergrounding distribution networks (0.4%) or randomly undergrounding transmission network (0.1%), demonstrating a superior cost-efficiency of the greedy strategy. The topological law for power outages that we found here may help local governments to plan resilience-based power networks in a changing climate.

**Discussion**

The results of this analysis demonstrate how dramatically the impact of hurricanes may increase over time, due to compound effects of storm and heatwave climatology change. For Harris County, the expected percent of customers experiencing at least one longer-than-5-day heatwave without power post-hurricanes in a 20-year period would increase alarmingly from 0.8% in the historical climate (1980-1999) to 15.5% towards the end of the 21st century (2080-2099), under the high-emissions scenario RCP8.5. For the current (2000-2019), near-future (2020-2039), and mid-of-the century (2040-2059) time frames, which are also highly relevant to decision making, this impact percentage would increase to 2.0%, 4.7%, and 6.2%, respectively (see Supplementary

Fig. S3). Even if we account for the uncertainty in the prediction of storm frequency [54] and remove the predicted increase in the storm frequency for the study area (by applying the frequency in the historical climate), the impact percentage would still increase significantly, to 9.5% towards the end of the 21st century (Supplementary Fig. S4). Only if we assume that the warming is well controlled under the low-emissions scenario RCP 2.6 and the hurricane activity (including both frequency and intensity) remains the same as the historical level, will the compound hazard change slightly, with the impact percentage changing from 0.8% in the historical climate to 1.0% towards the end of the century (Supplementary Fig. S5). If the RCP8.5 scenario is considered an upper bound [40] and the ideal RCP2.6 scenario a possible lower bound, this additional analysis indicates that the compound hazard risk may be largely avoided under rigorous climate mitigation policies in line with the Paris Agreement. However, there exist much uncertainties in the economic and political systems [40] as well as in the climate system (including possibly complex physical interactions between hurricanes and heatwaves); climate adaptation and risk mitigation are still necessary and urgent.

Other uncertainties exist in the modeling of the future compound hazard and risk. As the first attempt in quantifying the compound hazard risk, here we focus on the dominate power damage effects -- winds and induced falling trees [41]. Located relatively high above the sea level (see elevation map in Supplementary Fig. S1), our study area, Harris County, is affected mainly by extreme winds, as evident in historical events including Hurricanes Ike and Harvey (though lower Houston beyond the Harris County was heavily flooded by Harvey's extreme rainfalls and compound flood). Accounting for significantly less impact (< 10% hurricane-induces power outages; [41]), however, flooding and associated debris during hurricanes can also cause damage to the power system, and they usually impede the early recovery of a power network. Flood-prone regions may experience higher risks than estimated here, and future development of the modeling method must take flood impact into account. A precise prediction of flood-induced power outage and recovery requires accurate prediction of the magnitude and timing of the flooding (from storm surge and/or heavy rainfall) [42], power system vulnerability under flooding [43], and operational logistics of local utilities when repairing wetted power system components.

The scaling between the power network and the population in a naturally growing city may enlarge the affected population in the future, as power facilities usually develop much slower than the growth of local population (a 0.87-power scaling [44] Given this scaling effect, unbalances between local population and network density may increase, and thus future risks may be higher than those estimated herein. However, we do not account for the benefits of backup generators or solar panels here. These local devices could temporarily support residents who lose power from the main grid -- another possible way to mitigate impacts of hurricane-blackout-heatwave compound hazards. The power demand and dependence may reduce during hurricanes thanks to evacuation, but it may also increase if electrical vehicles are increasingly adopted and used for evacuation [45]. Also, as the temperature increases in the future climate, it is possible that utilities will have the incentive to upgrade the power system capacity to match with the higher temperature-related power demand. However, improving the power capacity would not significantly improve the power system reliability and resilience under hurricanes, even when the capacity is increased by 50% (see Supplementary Fig. S6).

Strategically undergrounding local distribution networks can efficiently enhance the resilience of the power system to adapt to climate change. As the power outages under hurricanes and daily operation are both dominated by the generalized scaling law (top 20% of local damages would trigger near 80% of total outage), the reliability-based enhancement of power grids against hurricanes can be considered jointly with daily-operation-based enhancement. This finding points further to the potential to develop a unified design framework for enhancing the power system resilience against various damage sources. The potential co-benefit and improved cost efficiency induced by the unified strategy might arouse more interest for local governments to mitigate the compound hazard risk. Furthermore, the power outage and compound hazard we consider herein can significantly disrupt local business and supply chains, leading to secondary losses [46], and the enhanced connectivity of local and global economics potentially would further foster the impact [47]. The coupled modeling of the compound hazard and induced economic disruption may be applied in future studies to quantify the cost-benefit [48] of proposed risk mitigation strategies.

To develop efficient risk mitigation strategies, quantifying the reliability and resilience of infrastructure systems under the impact of future compound hazards is essential but challenging. Risk analysis requires continuous updates wherein improved modeling and new data become available. As extreme climate events become more frequent, coastal megacities also develop rapidly [1, 49]. To ensure sustainable development, effective strategies to mitigate the hurricane-blackout-heatwave compound hazard, a pronounced example of physical-social connected extremes [50], should be carefully developed. This study demonstrates an interdisciplinary approach that integrates the-state-of-art climate science and infrastructure engineering to project climate change impact and develop risk mitigation strategies, with the goal of achieving resilient and sustainable communities.

## Materials and Methods

### Hurricane Projection.

TCs cannot be well resolved in typical climate models due to their relatively small scales, except perhaps in a few recently-developed high-resolution climate models (e.g., 32). Dynamic downscaling methods can be used to better resolve TCs in climate-model projections (e.g., 51), but most of these methods are computationally too expensive to be directly applied to risk analysis. An effective approach is to generate large numbers of synthetic TCs under reanalysis or GCM–projected climate conditions to drive hazard modeling. In this study, we apply large datasets of synthetic storms generated by a deterministic-statistical hurricane model [27]. This model uses thermodynamic and kinematic statistics of the atmosphere and ocean derived from observations or a climate model to produce synthetic TCs (hurricanes); it has been widely used to study hurricane wind, storm surge, and rainfall hazards (e.g., 28,52).

Specifically, we apply the synthetic hurricane datasets generated with this model by ref. [28] for the Houston area. Each synthetic storm passes within 300 km of Houston, with a maximum wind speed of at least 22 m/s (sensitivity analysis shows that the hazard modeling results are not sensitive to the storm selection radius when it is greater than 200 km). The datasets include 2000 synthetic hurricanes under the historical climate over the period of 1981–2000 based on the National Centers for Environmental Prediction (NCEP) reanalysis as well as 2000 synthetic hurricanes for the historical climate (1981-2000) and 2000 synthetic hurricanes for the future climate (2081-2100 under the high-emissions scenario RCP8.5) based on each of 6 CMIP5 GCMs (chosen based on data availability and following previous studies): the National Center for Atmospheric Research CCSM4, the United Kingdom Meteorological Office Hadley Center HadGEM2-ES, the Institute Pierre Simon Laplace CM5A-LR, the Japan Agency for Marine-Earth Science and Technology MIROC-5, GFDL-CM3.0 (NOAA Geophysical Fluid Dynamics Laboratory), and the Japan Meteorological Institute MRI-CGCM3.

To account for possible biases in the climate projection, we bias-correct storm frequency and landfall intensity and apply stochastic modeling to resample the storms. Specifically, for each GCM-driven projection, we bias-correct the projected storm frequency for the future climate by multiplying it by the ratio of the NCEP estimated frequency (calibrated to be 1.5 times/year for Houston using historical data [54]) and GCM estimated frequency for the historical climate, assuming no change in the model bias over the projection period, following refs. [53,54]. We apply the same assumption to bias-correct the projected landfall intensity (maximum wind speed) for the future climate, through cumulative density function (CDF) quantile mapping based on comparison of the NCEP and GCM estimated CDF of the annual maximum landfall intensity for the historical climate, similar to ref. [55], and reweighting each hurricane simulation based on the bias-corrected intensity distribution. To obtain a single projection for the future climate, we combine the bias-corrected projections from the 6 GCMs, weighted (as in ref. 56) according to their performance in estimating the CDF of the annual maximum landfall intensity for the historical climate compared to NCEP estimates. Finally, we stochastically resample the storms from the combined projection, assuming that storms arrive as a stationary compound Poisson process for the specific climate condition (ref. [54]); 10,000 20-year simulations are generated for the historical climate and 10,000 20-year simulations are generated for the future climate. The storm arrival times are matched with the heatwave analysis for the study area. For each sampled storm, we generate the spatial-temporal wind field, employing the classical Holland wind profile [57] and accounting for the effects of surface friction and large-scale background wind based on ref. [58], to drive the power grid failure analysis.

**Heatwave Projection.** The HI is calculated as a function of near-surface (at 2 m) air temperature, specific humidity, and surface pressure [24]. To be consistent with the hurricane simulation, we obtain these data (daily) for Houston from the NCEP reanalysis and 6 GCMs mentioned above, matched in time with the landfall (or at the closest point to Houston) of each generated synthetic storm. The GCM-projected future near-surface temperature is bias-corrected by adding to it the difference between the NCEP reanalysis and GCM-estimated historical temperature (monthly average cubic spline interpolated to daily). When combining the datasets of storm and heat events, we account for their possible correlation. Based on observations, ref. [24] found that TCs arrive after an anomalously high HI from amplified air temperatures and specific humidity; after TC passage, HI anomalies decrease to negative and return to zero within approximately 10 days. Ref. [24] neglected the interdependence between storms and heatwaves when estimating compound TC-heat events on a 30-day time scale. Here, concerning compound events on shorter time scales, we account for the interdependence statistically: we add the composite impact of hurricane passage to the meteorological variables used to calculate the HI, where the composite impact is estimated based on historical data (Fig. 3a in ref. [24]). Accounting for this correction reduces the HI values within 5 days of hurricane passage and thus the probability of compound hurricane-blackout-heat events defined in this study.

**Power System Modeling.** While various statistical models [7,59,60] have been developed to estimate hurricane-induced blackout, we employ a physics-based model to better account for future evolving factors, e.g., climate change, infrastructure upgrade, and utility maintenance. Specifically, we apply the power grid outage and recovery process model developed by ref. [29] to simulate hurricane impact on the electric power system in Harris County, TX. This system serves approximately 1.7 million customers in a service area around 4,600 km$^2$, and over 90% metropolitan households in West South America use air-conditioning [61]. The power grid includes high-voltage transmission networks, where 551 transmission lines connect 23 power plants and 394 substations, and low-voltage distribution networks, which contains ~40,000 branches [29] (Supplementary Fig. S1).

Given the hurricane wind hazard (i.e., local maximum wind during the storm; possible effect of wind dynamic is neglected given that the study area is relatively small), the power grid failure model first applies probabilistic fragility functions to estimate the damage to five main vulnerable component types of the power network: transmission substations, transmission lines, distribution nodes, distribution lines, and local distribution circuits. Component failures alter the power grid topology and may separate the power grid into unconnected subgrids. A direct current (DC)-based power flow simulation is performed to capture the power flow pattern in each subgrid, and the local demand is cut when overflow happens until the system achieves a steady state (refs. 32,33 used a similar approach). The power system is open and connects with systems outside the study area via transmission lines; the performance of the power grid outside the study area is assumed normal. The recovery model, developed based on emergency response plans and operational data, applies estimated recovery resources based on a priority-oriented strategy to repair damaged transmission substations, transmission lines, and critical facilities vital to public safety, health, and welfare before local distribution networks [29]. The power grid outage and recovery model was calibrated for the study area by ref [29] using observed data for Hurricane Ike (2008). We further evaluate the model adding data from Hurricane Harvey (2017). The same wind field modeling method applied to the synthetic storms is used for these two historical storms with storm characteristics (i.e., track, intensity, and size) taken from the extended best track data [62].

**Network Analysis and Enhancement Investigation.** After investigating the hurricane-blackout-heatwave compound hazard, we apply network analysis to investigate how the spatial pattern of power outage is related to the network pattern of local power distribution sectors, using our large synthetic hurricane dataset. Specifically, we investigate the generalized scaling relationship between the probability of local failures and their impact on the global outage, proposed by ref. [7], to understand the reliability of the power system. We analyze the connection between power outage and local distribution network topology by linking the power outage rate to the mean length of local distribution sectors. These analyses support the aim of designing efficient hazard mitigation strategies.

Various strategies have been proposed to enhance the post-hurricane resilience of power networks [36], e.g., adding recovery resources, applying stricter structural criteria, and undergrounding network branches. Adding significant recovery recourses or applying stricter design criteria would raise daily costs, while benefiting emergency recovery more than daily operation [34]. Thus, based on our findings, we design an undergrounding plan for Harris County to enhance the resilience of the power system. Specifically, we propose greedily reducing the mean length of local distribution networks by protecting a small portion of wires close to root nodes of the distribution networks. We compare this strategy with undergrounding strategies that randomly bury transmission lines and distribution sectors (similar to the generally-used uniform undergrounding strategies) to evaluate its efficiency in reducing future risks of hurricane-blackout-heatwave compound hazards.